# Diffuse Interplanetary Radio Emission: Shock Emission or a type III storm?

*Nat Gopalswamy, Sachiko Akiyama, Pertti Mäkelä, and Seiji Yashiro*

*Abstract* – We present a clear case of a Diffuse Interplanetary Radio Emission (DIRE) event observed during 2002 March 11-12 in association with a fast coronal mass ejection (CME). In the previous event reported [1], there were two CMEs, and a detailed analysis was required to pin down the underlying CME. In the event presented here, the CME association is unambiguous, and the DIRE is found to originate from the flanks of the CME-driven shock. We also provide quantitative explanation for not observing radio emission from the shock nose. We also clarify that DIRE is not a type III storm because the latter occurs outside of solar eruptions.

## 1. Introduction

Solar type II radio bursts are caused by ~10 keV electrons accelerated at shocks driven by coronal mass ejections. The accelerated electrons form an unstable beam-plasma system resulting in the production of Langmuir waves, which get converted into electromagnetic emission observed as type II bursts. Type II bursts have many variants as inferred from the radio dynamic spectra, which are plots of the radio intensity as a function of frequency of emission and time [2]. Type II bursts appear as slowly drifting features in the dynamic spectra and have starting and ending frequencies at various spectral domains: metric (m, >15 MHz), decameter-hectometric (DH, 1-15 MHz), and kilometric (km, <1 MHz). In terms of heliocentric distance, these domains correspond to <2 Rs, 2-10 Rs, and >10 Rs, respectively. The DH spectral window became available after the advent of the Radio and Plasma Wave (WAVES) instrument on board Wind [3]. The Diffuse Interplanetary Radio Emission (DIRE) emission reported recently [1] is one of the findings enabled by the DH spectral window and the coronagraph coverage [4] of the spatial domain in which this emission occurs. DIRE is so named because it is observed in the Wind/WAVES spectral range, but it is not clear whether it has a coronal counterpart. The DIRE is characterized by a series of short duration bursts within an envelope that drifts like type II radio bursts. The appearance is distinct from ordinary type II bursts, which do not have such fine structures. Another important property is the section of the shock front the emission originates from: at the flanks where the shock crosses streamer stalks. Ordinary type II bursts can also originate from the flanks as a continuation of the m type II bursts; such bursts do not require interaction with a streamer. Here we report on the 2002 March 11-12 event that can be unambiguously associated with a CME observed by the Large Angle and Spectrometric Coronagraph [4] on board the Solar and Heliospheric Observatory (SOHO).

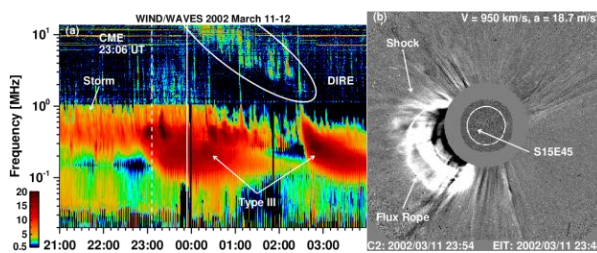

**Figure 1.** (a) Wind/WAVES dynamic spectrum showing the DIRE event encircled. The radio intensity is the excess over the background in dB (see color scale). Red and blue represent the highest and lowest intensities, respectively. A type III storm in progress and a group of type III burst are also seen in the dynamic spectrum. (b) SOHO/LASCO difference image at 23:54 UT showing the associated CME above the east limb with flux rope and shock parts marked. The solar source is a complex region at S15E45. The time of the CME in (b) (23:54 UT) and the first appearance time of the CME in the LASCO field of view (23:06 UT) are marked by the solid and dashed vertical lines in the dynamic spectrum.

## 2. The Solar Eruption Associated with DIRE

The Wind/WAVES radio dynamic spectrum showing the 2002 March 11-12 DIRE is presented in Figure 1 along with the associated CME in the SOHO/LASCO field of view (FOV). The DIRE starts at ~00:03 UT on March

12, very close to the time of the LASCO image in Figure 1. The starting frequency (~14 MHz) is just the upper edge of the WAVES dynamic spectrum, which means the DIRE is likely to have started at a slightly higher frequency. However, radio data obtained by the Hiraiso radio spectrograph do not show any corresponding feature down to 25 MHz (https://sunbase.nict.go.jp/solar/denpa/hirasDB/2002/03/020312a.gif). It is possible that the DIRE started somewhere between 14 and 25 MHz. The DIRE consists of a series of short-duration bursts that have a bandwidth of ~6 MHz. The emission ends around 02:18 UT on March 12 at ~2.4 MHz. The overall envelope of the DIRE drifted to lower frequencies at the rate of ~$1.6\times10^{-3}$ MHz/s. This value is within the range of drift rates obtained for a large number of type II bursts at DH wavelengths [5-6]. There is no fundamental-harmonic structure in DIRE, unlike the previous event [1]. While the drift rate of DIRE resembles that of a normal type II burst, the spectral features are quite different. The DIRE is also spectrally distinct from the ongoing type III storm noted in the dynamic spectrum.

Figure 1 also shows the associated white-light CME observed just a few minutes before the DIRE starts at 14 MHz. The CME is morphologically well defined with a flux rope structure and a shock-like structure clear in the northern flank. The CME first appeared in the LASCO/C2 FOV at 23:06 UT on March 11, barely showing up above occulting disk at 2.5 Rs. The CME LE is already at 4.69 Rs when the DIRE appeared at 14 MHz, and at 15.96 Rs when DIRE ended at 02:18 UT on March 12.

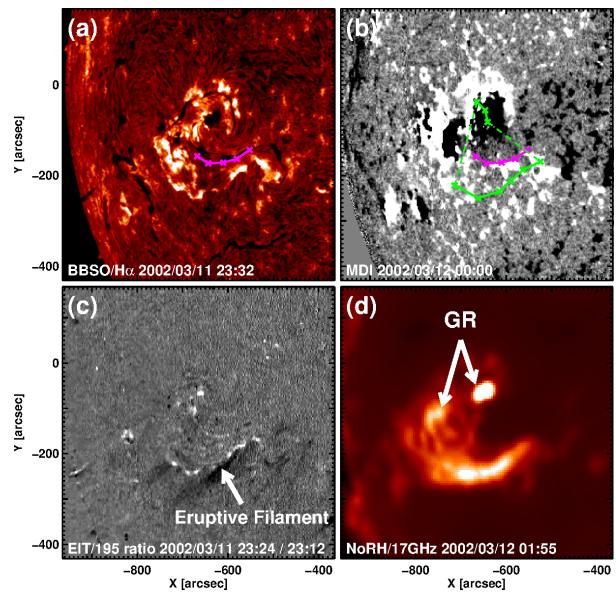

**Figure 2.** Solar source of the 2002 March 11 CME associated with the DIRE. (a) H-alpha image from the Big Bear Solar Observatory showing the neutral line (magenta) in question. (b) MDI magnetogram showing the complex source region with two active regions and an extended magnetic region. The neutral line of interest was located in the extended magnetic region. The green solid curves denote the feet of the post-eruption arcade (PEA). The dashed lines connect the end points to determine the reconnected flux. (c) Eruptive filament and the formation of footpoint kernels of the PEA. (d) The full extent of the PEA in a 17 GHz radio image obtained by NoRH. The two bright compact sources are gyro-resonance (GR) emission from sunspots. The brighter one is from AR 9866, while the weaker one is from AR 9870.

The CME originated from a filament eruption at S15E45 in a complex magnetic region consisting two NOAA active regions (ARs) 9866 and 9870 along with an extended region to the southwest. The filament erupted from a horizontal neutral line in the extended region (see Figure 2). The post eruption arcade (PEA) was complex, starting to the south of the filament location and connecting to the negative polarity region in AR 9866. The full extent of the PEA can be seen in the microwave (17 GHz) images obtained by the Nobeyama Radioheliograph (NoRH [7]) as in the snapshot shown in Figure 2. The foot points of the PEA traced from the SOHO's Extreme-Ultraviolet Imaging Telescope (EIT [8]) image are superposed on a Michelson Doppler Imager (MDI [9]) magnetogram showing complex connectivity. The total reconnected flux [10] computed from the PEA area ($1.38\times10^{20}$ cm$^2$) and the average magnetic field strength within the area (83 G) is $5.7\times10^{21}$ Mx.

There was a C-class soft X-ray flare listed in the online Solar Geophysical Data, however, there are contributions to the X-ray emission from other regions on the Sun, so we do not know the true time evolution of the solar source in X-rays. Fortunately, NoRH imaged the entire eruption at 17 GHz. The 17-GHz brightness temperature (Tb) averaged over the eruption region is shown in Figure 3. The flare in microwaves is very gradual, reaching a flat peak at ~01:20 UT on March 12.

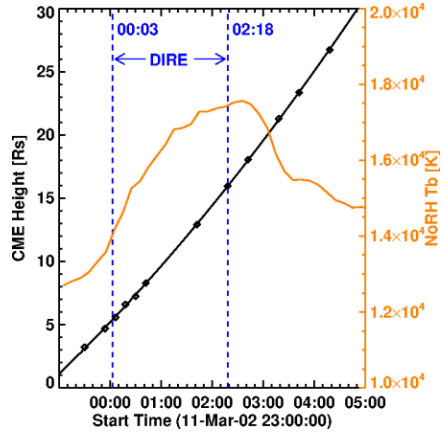

**Figure 3.** Height (h in Rs) - time (t in hours from 23:00 UT) measurements of the CME LE at position angle 118⁰ and the second-order polynomial fit (h = 1.09 + 3.92t + 0.17t$^2$). The DIRE start (00:03 UT) and end (02:18 UT) times are marked by the vertical blue dashed lines. Also shown is the 17 GHz brightness temperature averaged over the eruption region (right Y-axis).

Figure 3 also shows the height-time history of the CME LE, which clearly continues to accelerate with an average acceleration of ~18.7 m s$^{-2}$. Within the coronagraph FOV, the CME has an average speed of ~950 km s$^{-1}$. The initial acceleration of the CME is typically much higher and can be computed from the flare rise time and the average CME speed [11]. Using the 2.33-hr rise time of the 17 GHz profile in Figure 3 and the average speed of 950 km/s, we get the initial acceleration as ~113 m s$^{-2}$. The height – time measurement was made in the sky plane. We use a simple geometric deprojection by the 45⁰ angle the CME nose makes with the sky plane, to get the deprojected acceleration is ~ 158 m s$^{-2}$. Initial acceleration in the range of hundreds of m/s is characteristic of CMEs associated with filament eruptions from outside active regions [12]. In such events, the initial speed is typically much smaller than the final speed within the coronagraph FOV [13]. When the CME first appeared in the LASCO/C2 FOV at 23:06 UT its leading edge (LE) was at 2.5 Rs. In the next frame at 23:30 UT, the height was 3.2 Rs, giving an initial speed of 338 km/s (474 km/s deprojected). The last two data points in Figure 3 correspond to heights of 23.35 Rs and 26.73 Rs at 03:42 UT and 04:18 UT, respectively. The final speed, 1089 km/s (1525 km/s, deprojected) is much larger than the initial speed.

The average deprojected speed (V) is commensurate with the reconnected flux (F) [10]: V = 394×F$^{0.67}$. Substituting F = ~ 5.7×10$^{21}$ Mx in this equation, we get V = 1265 km/s, which is similar to the measured average speed ~1340 km/s (projection corrected from 950 km/s). While the speed is high, the different initial and final speeds within the coronagraph FOV have important implications for the ability of the CME to drive a shock and accelerate particles.

## 2.1 Where Does DIRE Originate from?

At the DIRE starting (00:03 UT) and end (02:18 UT) times, the CME LE (nose) is at of 5.36 Rs and 15.96 Rs, respectively (see Figure 3). When a simple geometrical deprojection is applied, these heights become ~7.51 Rs and 22.57 Rs, respectively. If the DIRE originates from the CME nose region, the local plasma frequency immediately ahead of the CME LE needs to be 14 MHz and 2.4 MHz at the two heights, assuming fundamental plasma emission. These frequencies correspond to electron densities of 2.4×10$^6$ cm$^{-3}$ and of 7.1×10$^4$ cm$^{-3}$, respectively. Even if the emission occurs at the harmonic of the local plasma frequency, the expected plasma frequencies are 7 MHz and 1.2 MHz. We can estimate the plasma frequencies based on the frequency of the plasma line at the Wind spacecraft. In the interplanetary medium, the density falls off as the square of the heliocentric distance (*r*). The plasma frequency, therefore, falls off as 1/*r*. Assuming that the plasma density distribution does not change significantly above the active region over the next several days, the measured plasma frequency (30 kHz) at Wind on 2002 March 16 is extrapolated to the heights 7.51 and 22.57 as 0.86 MHz and 0.29 MHz, respectively. Clearly these plasma frequencies are much smaller than the DIRE frequencies of 14 and 2.4 MHz. Therefore, we can rule out the possibility of DIRE originating from the CME nose region.

The curved shock in front of the CME flux rope cuts plasma levels of progressively increasing density as one goes away from the nose in the lateral direction (see Figure 1b). According to the Leblanc et al. [14] density model normalized to the observed 1-AU density (~11 cm$^{-3}$ corresponding to 30 kHz), the 14 MHz plasma level occurs at a heliocentric distance of ~2 Rs. This height is just beneath the LASCO/C2 occulting disk. Presence of streamers at the flanks would raise the 14 MHz plasma level to a larger height depending on the density enhancement in the streamers. Thus, we can conclude that the DIRE frequencies correspond to the flank region of the CME shock.

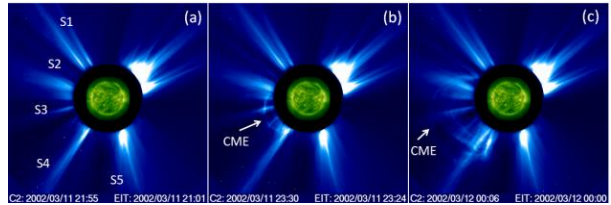

**Figure 4.** Several streamer structures above the east limb on either side of the CME in LASCO/C2 images. (a) A pre-event image at 21:55 UT with the streamers S1-S5 labeled. (b)The 23:30 UT image showing the CME in progress just before the DIRE onset. (c) The 00:06 UT image taken just after the DIRE onset. EUV images at 195 Å are superposed for context.

Figure 4 shows a set of LASCO/C2 images showing the coronal streamers S1-S5. As the CME moved out, streamers S2 and are clearly disturbed indicating interaction with the CME. On the other hand, streamers S1 and S5 do not show any changes. Although we do not know the exact streamer location with respect to the CME, it appears that the streamers are with the CME span along the line of sight. Another aspect of the streamer interaction worth mentioning is that the fast magnetosonic speed ($V_M$) is lower in streamers. When a weak shock enters a streamer, it becomes locally strong and accelerates electrons more efficiently.

## 2.2 Why Is the Shock Nose Radio Quiet?

From the height-time history in Figure 3 we see that the CME is relatively fast: ~950 km/s (~1340 km/s, deprojected). The reconnected flux in the source region is consistent with the observed speed and agrees with the statistical relation between the two parameters [7]. This speed is high enough to drive a shock and produce a type II burst, yet the CME nose was radio quiet. Whether a CME drives a shock or not depends on the relative importance of the CME speed and $V_M$ at a given height. It is well known that at a distance of ~3 Rs, $V_M$ peaks, falling off on either side of this distance [15-16]. The peak $V_M$ has been found to vary by a factor of 3 in the range ~500 km/s to ~1600 km/s [17]. The CME speed in the present case lies in this range, so it is possible that the nose did not drive a shock, or the shock was too weak to accelerate electrons. High $V_M$ in the corona results from a low density and high magnetic field strength. Such coronal regions are tenuous as indicated by the relatively dark regions in coronagraph images. The dim corona above the CME nose in Figure 4b is indicative of lower density and hence higher $V_M$ (assuming the ambient magnetic field does not vary much). Furthermore, the CME is accelerating through the coronagraph FOV, indicating that the early speed is likely to be small. The CME first appeared in the LASCO/C2 FOV at 23:06 UT with its nose at 2.5 Rs. In the next frame at 23:30, the nose height was 3.2 Rs, giving an early speed of 338 km/s, which deprojects to 474 km/s. This height range is near the $V_M$ peak (>500 km/s) and hence it is unlikely to drive a shock at the nose region. On the other hand, the flanks are at a lower height owing to the curved nature of the CME. For example, the flanks intersect with the occulting disk at 2 Rs, at a position angle ~60⁰ from the nose (see Figure 1b). Assuming the radio emission originates from the flank at a height of 1.75 Rs, we expect a local $V_M$ of 400 km/s [15]. A nose speed of 827 km/s indicates a speed of 414 km/s at the flank ~60⁰ away. The deprojected flank speed is therefore 586 km/s, indicating a shock with a magnetosonic Mach number of ~1.5.

For limb events, one can readily obtain the electron density in the pre-event corona from polarized brightness images [1]. In the present case, the source is located at E45, so the corona above the east limb gives only an approximate situation of the corona surrounding the CME. A LASCO/C2 polarized brightness image was obtained just before the CME onset. Unfortunately, the density inversion did not converge in many position angles as shown in Figure 5. Nevertheless, we can see that in the position angle corresponding to the CME, the pre-event corona had consistently lower density compared to the streamers. In particular, the streamer S2 has a density at least 2.5 times larger than that in the nose region (we do not know the exact factor because of data gaps in several streamer position angles). There is a clear trend that the nose is at a lower plasma level than the flank regions where the streamers are located.

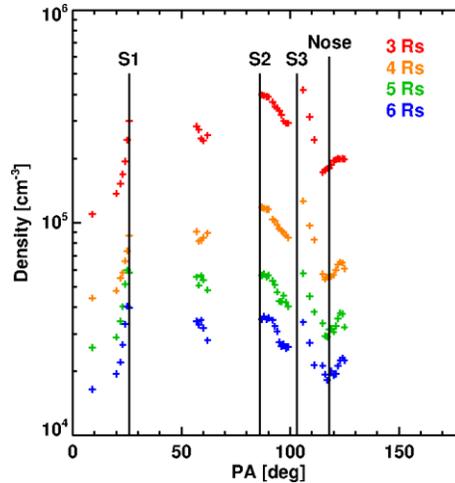

Figure 5. Coronal density as a function of the position angle (PA) at 4 heliocentric distances derived from the LASCO/C2 polarized brightness image at 22:12 UT on March 11, just before the eruption in question. The central PAs of streamers S1-S3 and the CME are indicated by the vertical lines. The density inversion did not converge at many PAs.

## 3  Discussion and Conclusions

We presented on a Wind/WAVES DIRE event associated with a fast halo CME. We showed by extrapolating the 1-AU electron plasma frequency to the corona that the plasma levels corresponding to the CME nose do not match the DIRE frequency. This is true irrespective of the emission mode (fundamental or harmonic of the plasma frequency). On the other hand, the flank regions are at lower heights and hence of higher plasma frequency allowing plasma emission. In addition to the plasma-level match, the magnetosonic speed is lower in the flank region, so even a lower flank CME speed is enough to make it super-magnetosonic.

The DIRE feature in the dynamic spectrum is somewhat similar to the "pure DH" or m-DH type II variants, however, these do not have the fine structure observed in DIRE. Although more investigation is needed, we can speculate that the pure-DH originates in the shock flank passing through a normal corona, while the DIRE originates in the shock flanks passing through nearby streamer stalks. Electron beams escaping along streamer stalks cause the short-duration, type III-like bursts in DIRE. Sometimes m-DH and DH-km bursts occur simultaneously at different frequencies: at a given time, the m-DH feature occurs at a higher frequency than the DH-km feature does [6]. These features are interpreted as emission coming from the nose (DH-km) and flanks (m-DH) [6]. In the event in hand, the nose emission is absent (weak shock or no shock) and the m-DH component is replaced by the DIRE. Such an explanation allows the possibility that the DIRE occurs along with nose (DH-km) emission in some events. In fact, such events are frequently observed especially in cases where the regular DH-km type II burst is very intense. Liu et al. [18] reported on a couple of such events, which they labeled "type III storm"; they did note that the envelope of the storm bursts drifted like a type II burst. We know that type III storms occur in active regions outside of solar eruptions. Therefore, "type III storms" identified by Liu et al. [18] are indeed DIRE events accompanied by type II emission from the nose region.